\documentclass[
aps,
prd,
reprint,
showpacs,
superscriptaddress,
longbibliography,
floatfix,
nobalancelastpage,
nofootinbib
]{revtex4-2}

\usepackage[bbgreekl]{mathbbol}
\usepackage{amsfonts}
\DeclareSymbolFontAlphabet{\mathbb}{AMSb}
\DeclareSymbolFontAlphabet{\mathbbl}{bbold}

\usepackage{linenoAMS}
\usepackage{amsmath}
\usepackage{comment}
\usepackage{siunitx}
\usepackage[colorlinks=true, citecolor=RoyalBlue,
linkcolor=BrickRed, urlcolor=ForestGreen]{hyperref}
\usepackage[all]{hypcap}

\usepackage{enumitem}
\usepackage{autofigs9}
\usepackage{braket}
\usepackage{comment}
\usepackage{mathtools}

\graphicspath{{figures/}}
\usepackage[usenames, dvipsnames, svgnames, table]{xcolor}
\usepackage[capitalise]{cleveref}

\usepackage{graphicx}
\usepackage{xfrac}
\usepackage{ulem}
\usepackage{cancel}

\definecolor{spring}{rgb}{0.7,0.9,0.7}
\definecolor{brick}{rgb}{0.7,0.2,0.1}
\definecolor{redHL}{rgb}{1.0,0.7,0.7}
\definecolor{blueHL}{rgb}{0.7,0.7,1.0}
\definecolor{ElectricBlue}{rgb}{0.49, 0.976, 1.0}
\definecolor{blueT}{rgb}{0.039, 0.729, 0.71}
\definecolor{DarkGreen}{rgb}{0, 0, 0}
\definecolor{blue}{rgb}{0.066, 0.310, 0.984}
\definecolor{blueT}{rgb}{0.039, 0.729, 0.71}
\definecolor{black}{rgb}{0, 0, 0}

\usepackage[capitalise]{cleveref}

\def\i{\mathrm{i}}

\newcommand{\VarWithSub}[2]{#1_{\mathrm{#2}}}

\def\Tin{\VarWithSub{T}{in}}
\def\lengthFC{\VarWithSub{L}{fc}}
\def\loss{\mathcal{L}}

\def\omegaFC{\VarWithSub{\omega}{fc}}
\def\OmegaSQL{\VarWithSub{\Omega}{SQL}}

\newcommand{\Qmat}[1]{\ensuremath{\mathbbl{#1}}}
\newcommand{\Qvec}[1]{\overset{\raisebox{-2pt}{\text{\tiny$\bm\rightarrow$}}}{\mathbbl{#1}}}

\newcommand{\Smat}[1]{\ensuremath{\mathbf{#1}}}
\newcommand{\Svec}[1]{\ensuremath{{\mathbf{#1}}}}

\newcommand{\QLO}{\ensuremath{\Qvec{v}^{\dagger}}}

\newcommand{\SLO}{\ensuremath{\Svec{v}^{\dagger}}}

\newcommand{\src}{\ensuremath{\text{s}}}
\newcommand{\Arm}{\ensuremath{\text{A}}}
\newcommand{\Src}{\ensuremath{\text{S}}}

\newcommand{\srm}{\ensuremath{\text{s}}}

\newcommand{\udarrow}{\ensuremath{{\uparrow}\hspace{-0.4ex}{\downarrow}}}
\newcommand{\aADF}{\ensuremath{a}}
\newcommand{\ADFtoU}{\ensuremath{\Svec{a}^{\uparrow}}}
\newcommand{\ADFtoL}{\ensuremath{\Svec{a}^{\downarrow}}}
\newcommand{\ADFto}{\ensuremath{\Svec{a}^{\udarrow}}}

\newcommand{\ADFU}{\ensuremath{\Svec{d}^{\uparrow}}}
\newcommand{\ADFL}{\ensuremath{\Svec{d}^{\downarrow}}}
\newcommand{\ADF}{\ensuremath{\Svec{d}^{\udarrow}}}
\newcommand{\ADFnorm}{\ensuremath{\delta}}

\newcommand{\eLOU}{\ensuremath{e^{\uparrow}}}
\newcommand{\eLOL}{\ensuremath{e^{\downarrow}}}
\newcommand{\eLO}{\ensuremath{e^
{\udarrow}}}
\newcommand{\eLOr}{\ensuremath{e'^{\udarrow}}}
\newcommand{\eLOno}{\ensuremath{\overline{e}^{\udarrow}}}
\newcommand{\eLOnor}{\ensuremath{\overline{e}^{\udarrow}}}

\newcommand{\eLOUs}{\ensuremath{e'^{\uparrow}_{s}}} 
\newcommand{\eLOUsnp}{\ensuremath{e^{\uparrow}_{s}}} 
\newcommand{\eLOLs}{\ensuremath{e'^{\downarrow}_{s}}}
\newcommand{\eLOUa}{\ensuremath{e'^{\uparrow}_{a}}}
\newcommand{\eLOLa}{\ensuremath{e'^{\downarrow}_{a}}}

\newcommand{\eLOUn}{\ensuremath{\overline{e}^{\uparrow}}} 
\newcommand{\eLOLn}{\ensuremath{\overline{e}^{\downarrow}}}
\newcommand{\eLOUsn}{\ensuremath{\overline{e}^{\uparrow}_{s}}} 
\newcommand{\eLOLsn}{\ensuremath{\overline{e}^{\downarrow}_{s}}}
\newcommand{\eLOUan}{\ensuremath{\overline{e}^{\uparrow}_{a}}}
\newcommand{\eLOLan}{\ensuremath{\overline{e}^{\downarrow}_{a}}}

\newcommand{\Hopo}{\ensuremath{\Smat{H}_{\text{O}}}}
\newcommand{\Hopor}{\ensuremath{\Smat{H}'_{\text{O}}}}

\newcommand{\SQZint}{\ensuremath{z}}
\newcommand{\SQZcosh}{\ensuremath{\cosh{\SQZint}}}
\newcommand{\SQZsinh}{\ensuremath{\sinh{\SQZint}}}

\newcommand{\SQZcosht}{\ensuremath{\cosh(2\SQZint)}}

\newcommand{\SQZcoshh}{\ensuremath{\cosh\Big(\frac{\SQZint}{2}\Big)}}
\newcommand{\SQZsinhh}{\ensuremath{\sinh\Big(\frac{\SQZint}{2}\Big)}}

\newcommand{\angPUMP}{\ensuremath{\psi}}
\newcommand{\angPUMPh}{\ensuremath{\psi}}
\newcommand{\angSQZ}{\ensuremath{\phi}}
\newcommand{\angADF}{\ensuremath{\Phi}}
\newcommand{\angLO}{\ensuremath{\zeta}}
\newcommand{\angDET}{\ensuremath{\Delta}}

\newcommand{\mm}{\ensuremath{\eta_{\text{fc}}}}
\newcommand{\Hr}{\ensuremath{\Qmat{H}_{\text{R}}}}
\newcommand{\Hfc}{\ensuremath{\Qmat{H}_{\text{fc}}}}
\newcommand{\Hifo}{\ensuremath{\Qmat{H}_{\text{IFO}}}}
\newcommand{\rp}{\ensuremath{r_+}}
\newcommand{\rmin}{\ensuremath{r_-}}

\newcommand{\bA}{\ensuremath{\Smat{A}}}
\newcommand{\bAi}{\ensuremath{\Smat{A}^{-1}}}

\newcommand{\fdloss}{\ensuremath{\eta}}
\newcommand{\fdrot}{\ensuremath{\theta}}
\newcommand{\fddep}{\ensuremath{\Xi}}
\newcommand{\thcav}{\ensuremath{\mathbf{\Theta_{\textbf{cav}}}}}

\newcommand{\SQZdB}{\ensuremath{N_{\text{dB}}}}

\newcommand{\IFOk}{\ensuremath{\mathcal{K}}}
\newcommand{\IFOr}{\ensuremath{\mathfrak{r}}}
\newcommand{\IFOg}{\ensuremath{\gamma_A}}

\begin{document}
\title{Probing squeezing for gravitational-wave detectors with an audio-band field }

\author{Dhruva Ganapathy}
\email{dhruva96@mit.edu}
\affiliation{Massachusetts Institute of Technology, Cambridge, MA 02139, USA}
\author{Victoria Xu}
\affiliation{Massachusetts Institute of Technology, Cambridge, MA 02139, USA}
\author{Wenxuan Jia}
\affiliation{Massachusetts Institute of Technology, Cambridge, MA 02139, USA}
\author{Chris Whittle}
\affiliation{Massachusetts Institute of Technology, Cambridge, MA 02139, USA}
\author{Maggie Tse}
\affiliation{Massachusetts Institute of Technology, Cambridge, MA 02139, USA}
\author{Lisa Barsotti}
\affiliation{Massachusetts Institute of Technology, Cambridge, MA 02139, USA}
\author{Matthew Evans}
\affiliation{Massachusetts Institute of Technology, Cambridge, MA 02139, USA}
\author{Lee McCuller}  
\affiliation{Massachusetts Institute of Technology, Cambridge, MA 02139, USA}

\date{\today}

\begin{abstract}
 Squeezed vacuum states are now employed in gravitational-wave interferometric detectors, enhancing their sensitivity and thus enabling richer astrophysical observations. In future observing runs, the detectors will incorporate a filter cavity to suppress quantum radiation pressure noise using frequency-dependent squeezing. Interferometers employing internal and external cavities decohere and degrade squeezing in complex new ways, which must be studied to achieve increasingly ambitious noise goals. This paper introduces an audio diagnostic field (ADF) to quickly and accurately characterize the frequency-dependent response and the transient perturbations of resonant optical systems to squeezed states. This analysis enables audio field injections to become a powerful tool to witness and optimize interactions such as inter-cavity mode matching within gravitational-wave instruments. To demonstrate, we present experimental results from using the audio field to characterize a 16 m prototype filter cavity.
\end{abstract}

\maketitle

\section{Introduction}
Quantum noise has become a limiting noise source across nearly the entire detection band of gravitational-wave instruments \cite{Barsotti_2018}, such as Advanced LIGO \cite{Buikema-PRD20-SensitivityPerformance}. During their last observing run, O3, LIGO and Virgo have demonstrated that quantum noise can be reduced by 3 dB above 100 Hz \cite{Tse-PRL19-QuantumEnhancedAdvanced, Acernese-PRL2019-squeezing}. Moreover, the GEO 600 interferometer has recently demonstrated a reduction of quantum noise by up to 6 dB at kilohertz frequencies \cite{GEO6dB_PhysRevLett.126.041102}. Still, future gravitational-wave observatories will require significantly higher levels of squeezing to realize the heightened sensitivities offered by order-of-magnitude longer interferometer baselines. In particular, emerging plans for next-generation facilities anticipate reaching 10 dB of broadband quantum noise reduction across the low-frequency audio band relevant to terrestrial gravitational-wave detection \cite{evanshorizon,ET}. 


%

The level of usable squeezing for reducing quantum noise is limited by optical losses and phase noise along the squeezed beam path\cite{Schnabel-PR17-SqueezedStates}.
A major obstacle to observing high squeezing levels is the difficulty of accessing squeezing degradation mechanisms within optical systems as complex as gravitational-wave interferometers. A compelling milestone in the development of squeezing was the mitigation of degradations through coherent control of squeezing \cite{Vahlbruch-PRL06-CoherentControl, Chelkowski-PRA07-CoherentControl}. Here, we extend those methods to provide \textit{coherent diagnostics} of squeezing. Traditionally, characterizations of the squeezed vacuum source have relied on switching operation modes to inject a bright ``seed'' or probe field at the carrier frequency \cite{Lam-JOBQSO99-OptimizationTransfer, Bowen-03-ExperimentsQuantum, Zhang-PRA03-QuantumTeleportation, Aoki-OEO06-Squeezing946nm, Takeno-OEO07-ObservationDB} to measure a nonlinear gain parameter. This nonlinear gain is then related to the squeezer's improvement of quantum noise as measured from the noise spectrum of the detector's readout photodetector \cite{dwyer_thesis, emil_thesis, Stefszky_2012}. 
However, a bright resonant field cannot be used to monitor the squeezed vacuum source during regular interferometer operation. Moreover, relying on the readout photodetector spectrum to determine the overall squeezing performance is an approach which
requires significant experimental overhead to first characterize the interferometer's non-quantum noise spectrum, then long averaging times to resolve the quantum contribution below the statistical noise of spectrum estimation \cite{McCuller-PRD21-LIGOQuantum}. This method is slower than typical interferometer noise timescales, obscuring real-time noise sources that influence squeezing. Switching operating modes is also undesirable in full-time observatories.

Characterizing squeezing with these established approaches becomes more demanding with the addition of frequency-dependent squeezing implemented using acoustic-bandwidth optical filter cavities \cite{McCuller-PRL20-FrequencyDependentSqueezing, Zhao-PRL20-FrequencyDependentSqueezed}, as planned for Advanced LIGO and Advanced Virgo in the upcoming observing run \cite{Abbott-LRR20-ProspectsObserving}, O4.
A filter cavity rotates the squeezed quadrature to simultaneously enable photon shot noise reduction at high frequencies as well as quantum radiation pressure noise reduction at low frequencies \cite{Caves-PRL80-QRPN, Kimble-PRD01-ConversionConventional}, overall delivering a broadband reduction of quantum noise. 
Both interferometer cavities and filter cavities apply complicated frequency-dependent and time-drifting degradations \cite{Kwee-PRD14-DecoherenceDegradation, McCuller-PRD21-LIGOQuantum}.
Maintaining high levels of squeezing through the filter cavity while optimizing to drifting interferometer parameters motivates developing additional diagnostics for near real-time probes of squeezing propagation throughout the system.

In this work, we demonstrate a new method to rapidly probe in-situ squeezing noise through a quantum filter cavity by using homodyne detection of an auxiliary \textit{audio diagnostic field} that co-propagates with the squeezed light.
%
%
This paper is organized as follows. Section \ref{sec:adf-intro} describes the experimental setup and mathematical analysis of the signal produced by homodyne detection of the audio diagnostic field. Sections \ref{sec:sqz-source}-\ref{sec:opt-sys} derive the relation of this audio signal to mechanisms of squeezing degradation, and then to the frequency response of an optical system utilizing squeezing. Section \ref{sec:fc} specializes this analysis to quantum filter cavities, and presents experimental measurements of our 16-m filter cavity. 
Finally, Section \ref{sec:ifo} simulates audio field measurements for the planned integration of a 300-m filter cavity with the Advanced LIGO detectors \cite{AplusDesign}. 

\section{Audio Diagnostic Field}
\label{sec:adf-intro}

\begin{figure}[h]
    \centering
    \includegraphics[width=\linewidth]{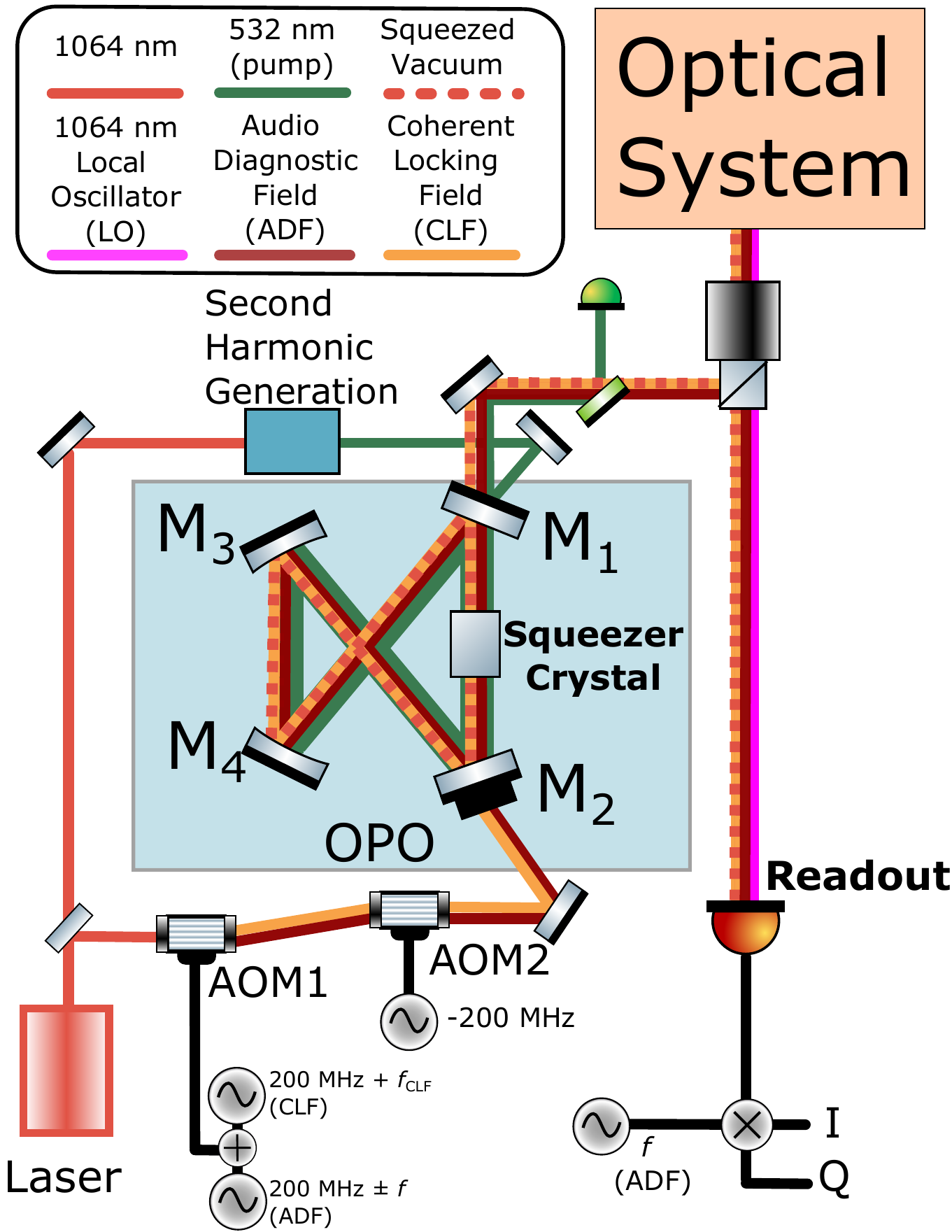}
    \caption{Experimental layout of a squeezer system with an audio diagnostic field. The output of the 1064 nm laser is split into two paths, one of which is upconverted to 532 nm by second harmonic generation, and used to pump the optical parametric oscillator (OPO). The other path of the 1064 nm laser is passed through a series of two acousto-optic modulators, AOM1 and AOM2, which generate both the ADF at a frequency $\pm f$, and the CLF at $f_{\text{CLF}}$; the CLF actuates on the AOM1 frequency $f_{\text{CLF}}$ to stabilize the squeezing angle. 
    The OPO is a doubly resonant bowtie cavity with mirrors $M_1, M_2, M_3, M_4$. The 532 nm OPO pump field is injected via $M_1$, while the 1064 nm CLF and ADF tones are injected into the OPO via $M_2$. 
    The audio diagnostic, coherent locking, and squeezed vacuum fields all exit the OPO via transmission through mirror $M_1$, and subsequently co-propagate through an optical system before homodyne detection with an external local oscillator (LO) field that has bypassed both the squeezer and optical system. The homodyne signal detected at the readout photodetector is demodulated at the audio frequency $f$ to obtain the real (I) and imaginary (Q) quadratures of the ADF-LO beatnote.  }
    \label{fig:figure_exp}
\end{figure}

Probing and controlling interactions of the squeezed vacuum field with the optical systems it propagates is difficult, because excess light at the main carrier frequency readily contaminates and degrades the level of observed squeezing. Therefore, the control of the squeezed field 
relies on off-resonant auxiliary coherent fields that co-propagate with the squeezed states from their source and through subsequent optical systems. 
The frequency-independent squeezed light source installed in the Advanced LIGO detectors \cite{Tse-PRL19-QuantumEnhancedAdvanced} already utilizes one such off-resonant auxiliary field, the coherent locking field (CLF) \cite{Vahlbruch-PRL06-CoherentControl}, for active phase noise control. 
However, to enable continuous squeezer operation without contaminating the astrophysical signal band, the CLF is typically detuned by a significant fraction of the linewidth of the optical parametric oscillator (OPO) ($\approx$9 MHz/20 MHz for our setup) that generates squeezing (see Fig \ref{fig:figure_exp}). As a result, the CLF is not representative of the squeezed carrier field itself, and does not directly sense the astrophysical signal band.

Here, we introduce a new auxiliary field which is generated at a small audio-frequency offset, well within the linewidth of the OPO and signal band. The small offset allows the transmitted audio field to experience the same transformations and degradations as the generated squeezed state. 

\subsection{Experimental Setup}
\label{sec:adf-expt}

Fig \ref{fig:figure_exp} contains the experimental setup of a squeezed vacuum source along with an injected audio sideband. Similarly to the LIGO squeezed vacuum source \cite{Tse-PRL19-QuantumEnhancedAdvanced}, in this experiment squeezed vacuum is produced by parametric down-conversion in a bowtie nonlinear OPO resonator \cite{Chua:11}, pumped by 532 nm light derived from the 1064 nm laser using second harmonic generation. The ADF is a single-frequency field that is shifted at acoustic frequencies from the 1064 nm carrier field, and it is generated using two acousto-optic modulators (AOMs). This allows for the creation of sidebands at arbitrary audio band frequencies and suppresses contamination by carrier frequency light\cite{Oelker-OO16-UltralowPhase}. The ADF is generated along with the CLF field and other auxiliary control beams. The ADF and the CLF are injected through the transmission port of OPO, and they co-propagate with the squeezed vacuum after exiting the OPO. 


Using the reflection of the CLF from the OPO, the CLF phase is stabilized with respect to the phase of the squeezer pump field using the AOM2 drive frequency as an actutator. As the ADF and CLF are generated simultaneously, the coherent control scheme also stabilizes the ADF phase with respect to the squeezer pump phase. The ADF and the squeezed vacuum field then beat with a local oscillator at the readout, generally after passing through an optical system. 

The beatnote between the ADF and the local oscillator is measured on a photodetector, where it is demodulated at the ADF sideband frequency into real (I) and imaginary (Q) quadratures. The nonlinear optical interaction of the OPO which generates squeezing also modifies the injected ADF.  As a result, the I and Q content carries information about the squeezing angle, squeezing level, local oscillator angle, and the optical system, all of which can be measured specifically at any chosen frequency. 

\subsection{Mathematical Description}
\label{sec:adf-theory}

The audio sideband field can be generated at a frequency offset $\pm f$ above or below the squeezed vacuum carrier field. In the sideband basis   , the upper and lower audio fields incident on the OPO are
%
\begin{align}
\ADFtoU &= 
          \aADF
    \begin{bmatrix}
     e^{-i {\angADF}}\\0
    \end{bmatrix}
    &
    \ADFtoL &=
              \aADF
    \begin{bmatrix}
      0\\
       e^{i {\angADF}} 
    \end{bmatrix}    ,
\end{align}
where $\aADF$ is the (real) sideband field amplitude, and $\angADF$ is a global phase of the audio field. Note that in these and future expressions, the ADF offset frequency $f$ that the injected amplitude, phase, and responses may depend upon, is implicit.   


After interacting with the OPO, the audio field emitted with the squeezing is
\begin{equation}  
   \ADF = \Hopo \ADFto ,
\end{equation}  
where $\Hopo$ is the transfer matrix from the ADF injection port to the transmission port of the OPO, derived for our cavity configuration in Appendix \ref{section:squeezer}. 
Because the audio sideband is generated at small detuning relative to the squeezer bandwidth, it can be treated as on-resonance in the OPO cavity. 

Injecting a single audio sideband, $\ADFto$, through the squeezer produces the output field, $\ADF$, occupying frequencies above and below the carrier
\begin{align}
  \ADFU &= \delta \begin{bmatrix}
    \alpha e^{-i\angADF}  \\ \beta e^{i(2\angPUMP-\angADF)}
  \end{bmatrix}
         &
  \ADFL &= \delta \begin{bmatrix}
    \beta e^{-i(2\angPUMP-\angADF)} \\ \alpha e^{i\angADF} 
  \end{bmatrix} ,
  \label{eq:ADF_simp}
\end{align}
with an overall scale factor 
\begin{equation}
  \ADFnorm = \aADF \dfrac{t_1t_2}{ r_1^2 -2 r_1\SQZcosh+1} 
  \label{eq:ADF_norm}
\end{equation}  
and relative sideband amplitudes
\begin{align}
  \alpha &= 1-r_1\SQZcosh
  & 
    \beta &=  r_1\SQZsinh 
\label{eq:alpha_beta}
\end{align}
that depend on the level of generated squeezing. Here, $\SQZint$ is the single-pass squeeze factor through the OPO's internal crystal, $\angPUMP$ is the phase of squeezer pump field, $t_1$ and $t_2$ are the transmissivities of mirrors $M_1$ and $M_2$, respectively, and $r_1$ is the reflectivity of $M_1$. Note that these parameters, and equations following from them, depend on the specific OPO design. Values for alternate OPO layouts are given in App. \ref{sec:alt_opo}.



Beating the transmitted audio field against a local oscillator field $\SLO$ with phase $\angLO$
%
%
\begin{equation}
  \SLO = \begin{bmatrix}
    -i e^{i\angLO}& i e^{-i\angLO}
  \end{bmatrix} / \sqrt{2}
\end{equation}        
produces the audio beatnote $\eLO$ in the photodetector readout of
\begin{equation}
     \eLO = \SLO \ADF.
\end{equation}    

Consider homodyne detection of the upper audio sideband after injection through the squeezer. This ADF-LO beatnote is given by
\begin{equation}
 \eLOU = -\frac{i}{\sqrt{2}}\ADFnorm {\left( \alpha e^{i\angSQZ}-\beta e^{-i\angSQZ} \right)},
     \label{eq:LO}
\end{equation}
here expressed as a function of the squeezing angle $\angSQZ$. Reducing this equation to a function of $\angSQZ$ is possible because we use the coherent control scheme employed in LIGO \cite{Tse-PRL19-QuantumEnhancedAdvanced}, which stabilizes the relative phase between the audio diagnostic, local oscillator, and squeezer pump fields.
As a result, the audio field is phase-stable with the pump field, setting $\angPUMPh = \angADF$. Coherent control also maintains the squeeze angle with respect to the pump field as $\angSQZ =  \angLO-\angPUMPh$. By our conventions, $\angSQZ = 0$ corresponds to squeezing (suppression of quantum shot noise) while $\angSQZ = \pi/2$ corresponds to anti-squeezing (amplification of quantum shot noise). 

The ADF-LO beatnote expression contains the factors $\delta$, $\alpha$, and $\beta$ which vary by the squeezing amplitude. Calibrating these factors and the overall magnitude of $\delta$ enables in-situ measurements of the squeezing parameters and intervening losses using the ADF beatnote.



\begin{figure}
    \centering
    \includegraphics[width = \linewidth]{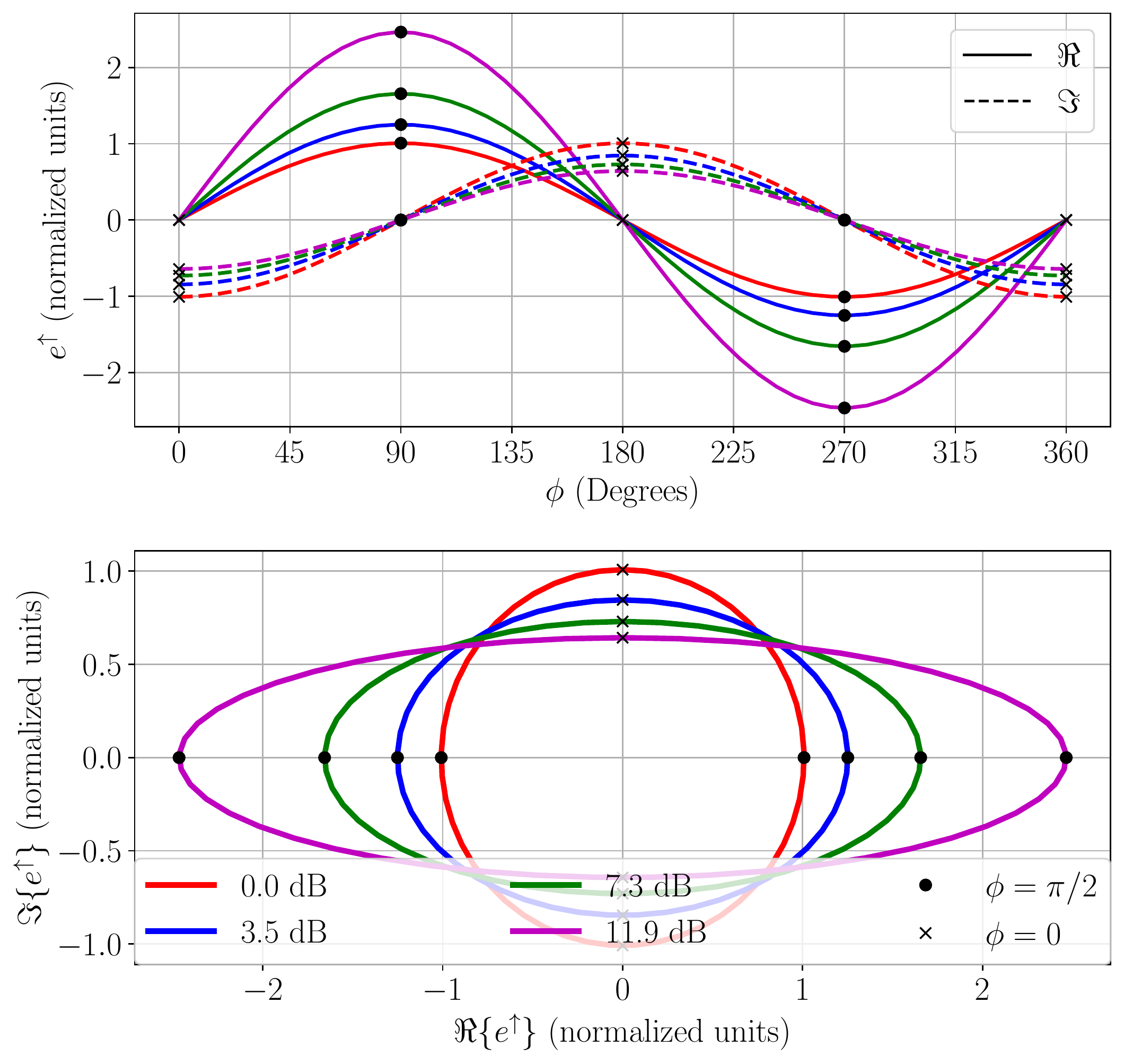}
    \caption{Demodulation space spanned by the ADF-LO signal $\eLOU$ for various squeezing levels. In both plots, the different colors correspond to different levels of generated external squeezing (expressed in dB of noise reduction; see Eq. \ref{eq:dB}) by an OPO with a reflectivity $r_1$ of 0.935. The upper plot shows the real ($\Re$) and imaginary parts ($\Im$) of the signal as a function of squeezing angle $\angSQZ$ in degrees. The crosses correspond to squeezing ($\angSQZ = 0$) while the dots correspond to anti-squeezing ($\angSQZ = \pi/2$). The lower shows a parametric plot of the real and imaginary quadratures of demodulated ADF-LO beatnote signal. Here, the dot and cross markers also correspond to the respective major and minor axes of the ellipse. Their ratio can be used to compute the external squeezing level generated by the OPO (Eq. \ref{eq:e_gain}). The units of $\eLOU$ have been chosen to normalize the case of no squeezing, $z=0$, to a unit circle.} 
    \label{fig:demod}
\end{figure}

\section{Characterising the Squeezed Vacuum Source}
\label{sec:sqz-source}
Before probing mechanisms which degrade the observable levels of squeezing, one can first characterise the squeezed vacuum state generated by the source. The audio field can provide an accurate and in-situ probe of the squeezed vacuum source due to its co-propagation and close detuning with the squeezed vacuum field. The level of squeezing generated at the output of the squeezer, and the angle of the squeezed state are key in calculating the maximum possible reduction in quantum noise that the squeezer can offer. 

To begin, the ADF-LO beatnote obtained in Eq. \ref{eq:LO} can be expressed in terms of squeezer parameters as
\begin{equation}
    \eLOU =\frac{\delta}{\sqrt{2}} \Big((1-e^{-\SQZint}r_1)\sin \angSQZ -i(1-e^{\SQZint}r_1)\cos \angSQZ\Big) ,
    \label{eq:e_ellipse}
\end{equation}
Demodulating the beatnote signal into real (I) and imaginary (Q) quadratures, one can parametrically plot the two quadrature signals as a function of squeezing angle $\angSQZ$ to produce the ellipse shown in Fig. \ref{fig:demod}.

\textbf{Squeezing angle.} The location of the signal on the ADF-LO ellipse corresponds to the squeezing angle $\angSQZ$
\begin{equation}
 \angSQZ = -\arctan{\Bigg(\frac{1}{G}\frac{\Re(\eLOU)}{\Im(\eLOU)}}\Bigg),
 \label{eq:SQZang}
\end{equation} 
where $G$ is the ratio between the largest (anti-squeezing) and smallest (squeezing) magnitude signals on the ellipse
\begin{equation}
    G = \frac{\alpha+\beta}{\alpha-\beta}=\frac{1-r_1e^{-\SQZint}}{1-r_1e^{\SQZint}} .
    \label{eq:GR}
\end{equation}
This quantity can be easily measured experimentally by rotating the squeezing angle $\angSQZ$ using the control system, while recording the minimum and maximum of the I and Q demodulation magnitude given by $|\eLOU|$. 

\textbf{Generated squeezing level.} The level of squeezing generated by the OPO at its output is related to the shape of the ADF-LO beatnote ellipse from the ratio $G$. 
Inverting Eq. \ref{eq:GR} allows one to calculate $\SQZint$ from a measurement of $G$
\begin{equation}
    z = \log\left(\dfrac{G-1+\sqrt{(G-1)^2+4Gr1^2}}{2Gr_1}\right).
\end{equation}
This internal squeezing level $\SQZint$ is then used to calculate parameters of fields transmitted through the squeezer, such as $\delta, \alpha$, and $\beta$ for the propagated audio field.

To calculate the total squeezing level generated by the source, which consists of the nonlinear crystal inside an optical resonator, the effective, or external, squeeze factor $Z$ is calculated to account for the resonator configuration. This effective squeeze factor is derived for our OPO cavity in Appendix \ref{section:NLG} to be
\begin{equation}
    Z = \log\Big(\frac{e^z - r_1}{1-r_1e^z}\Big).
    \label{eq:zZe}
\end{equation}
In terms of the ADF-LO beatnote ratio $G$, the effective squeeze factor is
\begin{equation}
     Z= \log\Bigg(\dfrac{G-1+\sqrt{(G-1)^2+4 G r_1^2}}{2r_1}\Bigg),
     \label{eq:e_gain}
\end{equation}
which corresponds to a quantum noise reduction in decibels of
\begin{equation}
    \SQZdB = 10\log_{10}(e^{-2Z}) = -8.6Z.
    \label{eq:dB}
\end{equation}
The factor $e^Z$ represents the (field) gain of the OPO as a parametric amplifier acting to squeeze the vacuum states at the carrier frequency. The ADF gain ratio $G$ is similar to, but not identical to, the conventional nonlinear gain of seeded carrier light. Appendix \ref{section:NLG} has more details.

\textbf{Ellipse normalization}. The ADF is transmitted from M2 to M1 of the OPO, while the squeezed field is obtained in reflection from M1. As a result, the OPO transforms the coherent ADF and squeezing ellipses differently. This distinction is notable when calculating external squeezing levels based on normalization of the ADF-LO beatnote ellipse. In brief, the ellipses differ because a field transmitting through the OPO can only be maximally squeezed by 6 dB in power (corresponding to a minimum normalized minor axis of 1/2) unlike the vacuum field reflected from the OPO, which ideally reaches arbitrary squeezing levels (sending the minor axis to 0).

Mathematically, this can be seen by calculating the minor axis magnitude (i.e., with $\angSQZ=0$) of the ADF-LO beatnote $\eLOUs$ in the limit of high and negligible squeezing, respectively:
\begin{align}
  \lim_{e^z \rightarrow r_1^{-1}}|\eLOUsnp| &=  \frac{a }{ \sqrt{2}}\frac{t_2}{t_1},
                                            &
  \lim_{e^z \rightarrow 1}|\eLOUsnp| &=  \frac{a }{ \sqrt{2}}\frac{t_1 t_2}{1 - r_1}.
\end{align}
The ratio of the two gives the high squeezing limit of the normalized minor axis of \cref{fig:demod}. As the OPO’s nonlinear gain nears the threshold of oscillation ($e^{-Z} \rightarrow 0$, or $e^z \rightarrow r_1^{-1}$), the maximal external squeezing level is reached; here, the normalized ADF minor axis reduces to a minimum value of $1/2$ when $1-r_1 \approx t_1^2/2$, while the major axis increases continuously as $\delta$ diverges.

This indicates that ADF measurements of $G$ are primarily determined by the major axis at high squeezing levels, and are not limited by the ability to resolve the minor axis above noise. The above formulas can also provide a convenient means of calibrating $a$ in the high and low squeezing limits. 

\textbf{Squeezing losses.} Loss, such as optical loss and photodetector loss, mixes coherent vacuum with squeezed vacuum. It is a dominant source of squeezing degradation. Losses reduce the ADF-LO beatnote magnitude $\eLO$ by a factor of $\sqrt{1 - \Lambda}$, where $\Lambda$ is the total fraction of squeezed vacuum that is replaced with coherent vacuum due to loss. With $\Lambda$, the ADF-LO beatnote from Eq. \ref{eq:LO} (setting $\phi=0$) is modified to 
\begin{equation}
     \eLOU = -i\ADFnorm {\left( \alpha-\beta\right)}\sqrt{1-\Lambda}.
     \label{eq:loss1}
\end{equation}

Calibrating the ADF signal to a known level of loss $\Lambda$ and a known level of squeezing, corresponding to fixed $\alpha$ and $\beta$, allows us to monitor losses as they drift over time. Using $\alpha'$, $\beta'$ and $\delta'$ to account for drifts in the generated squeezing level over time, the loss at a later time $\Lambda'$ is 
\begin{equation}
    \sqrt{1-\Lambda'} = \sqrt{1-\Lambda}\frac{\delta}{\delta'}\frac{\alpha-\beta}{\alpha'-\beta'}.
\end{equation}
%

\textbf{Squeezer phase noise estimation.} In addition to loss, phase noise, which refers to fluctuations in the squeezing angle \cite{Dwyer2013} is another important mechanism of squeezing degradation. These angle fluctuations mix a small portion of anti-squeezing into the squeezing quadrature, effectively decreasing the squeezing level observed at readout. As the amount of generated squeezing increases, the increase in quantum noise amplification by anti-squeezing is larger in magnitude than the suppression of quantum noise by squeezing. Physically, this results in a limit to how much squeezing can be generated before phase noise begins to reduce measured squeezing. 

The ADF provides a convenient way to experimentally measure squeezer phase noise, by way of measuring small angle fluctuations in the squeezing ellipse described through Eq. \ref{eq:e_ellipse}. Measuring angle fluctuations provides an estimate of the phase noise along the squeezing path. By setting the quadrature angle to squeezing ($\angSQZ = 0$), Eq. \ref{eq:SQZang} can be expanded in the limit of small angle fluctuations $\Delta\angSQZ\ll1$ to obtain an expression for the RMS squeezer phase noise,
\begin{equation}
    \Delta\angSQZ = \frac{1}{G}\sqrt{ 
    \bigg\langle \bigg(\frac{\Re(\eLOU)}{\Im(\eLOU)} \bigg)^2\bigg 
    \rangle}.
    \label{eq:phase_noise}
\end{equation}

Experimentally, this would involve rotating the ADF-LO beatnote signal completely into one demodulation quadrature, and measuring the RMS fluctuations in the orthogonal quadrature.

In the above section, we have shown that ADF provides a rapid and convenient way of making accurate in-situ measurements of the squeezer and time-varying system losses. Traditional methods of characterizing the squeezed vacuum source are based on injecting a strong field at the carrier frequency through the squeezer, which cannot be used during regular squeezing operation, or using photodetector quantum noise spectra, which requires long averaging times \cite{Tse-PRL19-QuantumEnhancedAdvanced}. Additionally, as loss and phase noise both effectively lower the levels of measured squeezing and anti-squeezing, it is hard to separate those effects using quantum noise spectra alone. 

\section{Characterizing Optical Systems}
\label{sec:opt-sys}

After characterizing the squeezed state source parameters, the audio diagnostic field can probe how squeezed states rotate, degrade, and dephase as they propagate through an optical system. Gravitational-wave interferometers are the key target of study using the ADF technique, and they have several properties that our study of squeezing must accommodate. First, interferometers typically read with a fixed local oscillator angle, $\zeta \approx 0$, to measure phase shifts of the light in their arms. Second, because of their high arm power, they affect optical states through quantum radiation pressure noise, which applies a frequency-dependent parametric gain to the squeezed state or fields incident on the interferometer.

Our analysis here builds upon the work in Ref.~\cite{McCuller-PRD21-LIGOQuantum}, which derives four squeezing parameters that together comprehensively describe the quantum noise spectrum observed in the Advanced LIGO detectors at the time of observing run O3. From Ref.~\cite{McCuller-PRD21-LIGOQuantum}, the four parameters are: rotation, $\fdrot(\Omega)$; dephasing, $\fddep(\Omega)$; readout-quadrature parametric gain, $\Gamma(\Omega)$, and efficiency, $\fdloss(\Omega)$. Each parameter is a function of the observation frequency in the spectrum, $\Omega/2\pi$, so this dependence is assumed in following expressions. These characteristics relate to the observed quantum noise spectrum $N$, normalized such that $N=1$ for shot noise\footnote{The approximation indicates that the loss term $1-\eta$ is not exact, as it depends on the location of the losses along the squeezing path. It is accurate when $\Gamma \sim 1$.}.
\begin{align}
  N
  &\approx \eta \Gamma \big(S_{\!{-}}\cos^2(\angSQZ {+} \theta) + S_{\!{+}} \sin^2(\angSQZ{+} \theta) \big) + 1 {-} \eta
    \label{eq:metric_N}
  \\
  S_{\!\pm} &\equiv (1 - \Xi)e^{\pm2Z} + \Xi e^{\mp2Z}
\end{align}
where $S_{-}$ and $S_{+}$ are expressions for the squeezed and anti-squeezed noise power, degraded by dephasing.


Let us assume that the optical system can be described by a generic complex 2$\times$2 matrix, $\Hr$, representing the linear frequency-dependent response of the system. This paper so far has used a sideband basis to show how the OPO transforms the upper and lower ADF optical frequencies. Here, $\Hr$ is instead in the quadrature basis, to represent the transformation of incoming and outgoing phase and amplitude quadratures at each frequency. Both representations are physically equivalent, but the quadrature basis is convenient given the formulas for radiation pressure effects and the fixed LO angle. 

Fully characterizing an arbitrary optical system requires both readout quadratures ($\angLO =0,\pi/2$) and both the upper and lower ADF signal injections to measure all of the terms of $\Hr$. Thus, one cannot fully characterize the optical response of interferometers due to the fixed readout quadrature; however, the ADF can entirely measure the parameters relevant to squeezing. As in Ref.~\cite{McCuller-PRD21-LIGOQuantum}, we combine the response of an optical system $\Hr$ and local oscillator $\QLO$ to obtain the homodyne observable with frequency-dependent quadratures, $m_q$ and $m_p$,
\begin{equation}
  \begin{bmatrix}
    m_q & m_p
  \end{bmatrix}
  \equiv \QLO \Hr .
  \label{eq:m_qp}
\end{equation}
To calculate the ADF-LO beatnote, the previously defined fields must first be expressed in the quadrature basis.
Using the formalism of \cite{Buonanno-PRD01-QuantumNoise,Kwee-PRD14-DecoherenceDegradation}\footnote{These references use different phasing sign conventions. This paper maintains consistency with Ref. \cite{McCuller-PRD21-LIGOQuantum}.}, the local oscillator field can be transformed with the following, 
\begin{equation}
  \QLO = \SLO \bAi = \begin{bmatrix} \sin(\angLO)& \cos(\angLO)\end{bmatrix},
\end{equation}
which uses the matrices 
\begin{align}
  \bA &= \frac{1}{\sqrt{2}}\begin{bmatrix} 1 & 1 \\ -i & i\end{bmatrix} ,
                                             &
  \bAi &= \frac{1}{\sqrt{2}}\begin{bmatrix} 1 & i \\ 1 & -i\end{bmatrix} 
\end{align}
to change between sideband/ladder and quadrature/Hermitian operator basis.

Then, $m_p$ and $m_q$ can be used to calculate the frequency-dependent loss $\fdloss(\Omega)$, rotation $\fdrot(\Omega)$ and dephasing $\fddep(\Omega)$ experienced by the squeezed state as it interacts with an optical system $\Hr$.
From Sec. IV and App. A of Ref. \cite{McCuller-PRD21-LIGOQuantum}, the squeezing parameters relate to the quadrature observables as\footnote{The expression for $\fdrot$ differs from the approximation of Eq. 37 in Ref. \cite{McCuller-PRD21-LIGOQuantum}. It is nearly numerically equivalent to the frequency-dependent rotation given by the singular value decomposition in App. A of Ref. \cite{McCuller-PRD21-LIGOQuantum}. Together, Eqs. \ref{eq:fdloss}-\ref{eq:fddep} obviate the need for a decomposition.} 
\begin{align}
    \fdloss\Gamma &= |m_p|^2+|m_q|^2 ,
    \label{eq:fdloss}
    \\
  \fdrot &= \frac{1}{2}\arg\left( \frac{m_p + i m_q}{m_p - i m_q}  \right) ,
    \label{eq:fdrot}
    \\
    \fddep &= \frac{1}{2}-\sqrt{\frac{(|m_p|^2-|m_q|^2)^2+4(\Re\{m_q m_p^*\})^2}{4(|m_p|^2+|m_q|^2)^2}} .
    \label{eq:fddep}
\end{align}
%

The ADF is driven and demodulated at a specific frequency $f$, and can measure $m_p$ and $m_q$ at $\Omega=2\pi f$.
Propagating the ADF through an optical system $\Hr$ modifies the ADF-LO beatnote from Eq. \ref{eq:LO} to
\begin{equation}
  \eLOr = \QLO \Hr \bA \ADF 
  \label{eq:eLO_unnorm}
\end{equation}
in the quadrature basis.
The following calculations assume a constant local oscillator angle $\zeta$. The control systems and our definition of the squeezing angle $\angSQZ$ make the squeezing and anti-squeezing conditions relative to the fixed LO angle, even if it is not exactly $\zeta=0$ as expected for perfect DC readout in an interferometer. 

The audio beatnote can be then calculated at squeezing ($\angSQZ=0$),
\begin{align}
  \eLOUs &= \frac{\delta}{\sqrt{2}}(m_q(\alpha + \beta) - i m_p(\alpha-\beta))\\
  \eLOLs &= \frac{\delta}{\sqrt{2}}(m_q(\alpha + \beta) + i m_p(\alpha-\beta))) 
\end{align}
and anti-squeezing ($\phi=\pi/2$),
\begin{align}
  \eLOUa &=\frac{\delta}{\sqrt{2}}(m_p(\alpha + \beta) + i m_q(\alpha-\beta))\\
  \eLOLa &= \frac{\delta}{\sqrt{2}}(m_p(\alpha + \beta) - i m_q(\alpha-\beta)) .
\end{align}
Rearranging the above equations, the squeezing parameters $m_p$ and $m_q$ can be obtained in terms of the measured ADF-LO beatnote,
%
%
\begin{equation}
 m_p = \frac{\eLOLa + \eLOUa}{\sqrt{2}\delta (\alpha + \beta)} = -\frac{\eLOUs - \eLOLs}{\sqrt{2}\delta i(\alpha - \beta)}  .
 \label{eq:mp_unno}
\end{equation}
\begin{equation}
m_q = \frac{\eLOUs + \eLOLs}{\sqrt{2}\delta(\alpha + \beta)} =  \frac{\eLOUa - \eLOLa}{\sqrt{2}\delta i(\alpha - \beta)} .
\label{eq:mq_unno}
\end{equation}

After passing the ADF through an unknown optical system, the ADF is modified and the ADF-LO beatnote ellipse (Sec. \ref{sec:sqz-source}) measurement of the level of generated squeezing becomes biased. Instead, these modified ADF-LO beatnote signals $\eLOr$ can be combined to make an unbiased measurement of $G$ that is applicable to any system using
\begin{align}
  G &= \frac{\alpha + \beta}{\alpha - \beta} = -i\frac{\eLOUa + \eLOLa}{\eLOUs - \eLOLs} = i\frac{\eLOUs + \eLOLs}{\eLOUa - \eLOLa},
  \label{eq:ratio}
\end{align}
which can then be put into Eq. \ref{eq:e_gain} to calculate the generated squeezing level. This generalizes Eq. \ref{eq:GR} for systems that may affect sideband balancing from nonlinear interactions, like radiation pressure.
Once $Z$ and $\SQZint$ are known, the overall gain $\delta$ can be determined to calibrate the ADF signal. The ability to combine upper and lower ADF measurements allows squeezer parameters to be characterized in-situ even for optical systems that may unbalance sidebands or apply parametric gain. This can enable long term in-situ study of the squeezer system stability, without invasive changes to the interferometer operating mode.

It is worthwhile here to point out the significance of the ``symmetrization'' implied by the sums and differences of \cref{eq:mp_unno} and \cref{eq:mq_unno}. The ADF, as proposed, is created by injecting a single upper or lower sideband into the OPO. At cost of increased complexity, one could alternatively inject balanced sidebands with relative phases chosen to create pure amplitude or phase quadrature modulations in the coherent field. Such injections would more directly measure $m_p$, $m_q$ in two separate measurements. The sums and differences above achieve the same goal, but avoid the experimental complexity of stably creating and phasing two audio field frequencies into AOM1.

\textbf{Normalized beatnote measurements.} The audio field can serve as an intermediary diagnostic for a single optical system embedded in a larger composite system. The response of an individual system can be isolated by normalizing the ADF-LO beatnote measurements between experimental configurations where the ADF does, or does not, pass through the optical system $\Hr$ using  
\begin{equation}
  \eLOnor = \frac{\QLO \Hr \bA \ADF}{\QLO \bA \ADF\sqrt{\eta_{\text{rel}}}} .
  \label{eq:eLO_norm}
\end{equation}   
Realistically, when the ADF is not passing through the optical system, it is redirected to a diagnostic readout that has a simple response including only the relative detection efficiency $\eta_{\text{rel}}$. That simple response is indicated in the denominator of \cref{eq:eLO_norm}. The normalized ADF-LO beatnote is calibrated to be unity on the diagnostic reference. The quadrature observables can then be calculated from the normalized ADF measurements as



\begin{equation}
 \frac{m_p}{\sqrt{\eta_{\text{rel}}}} = \frac{\eLOUsn + \eLOLsn}{2} = \frac{\eLOUan + \eLOLan}{2}   
 \label{eq:mp_no}
\end{equation}
\begin{equation} 
 \frac{m_q}{\sqrt{\eta_{\text{rel}}}}  = \frac{(\eLOUsn - \eLOLsn)}{2i}\Big(\frac{\alpha - \beta}{\alpha + \beta}\Big) = \frac{(\eLOUan - \eLOLan)}{2i}\Big(\frac{\alpha + \beta}{\alpha - \beta}\Big) .
 \label{eq:mq_no}
\end{equation}
Eq. \ref{eq:ratio} can also be re-written in terms of the normalized signals and used similarly to measure generated squeezing levels
\begin{align}
  G = \sqrt{\frac{(\eLOUsn - \eLOLsn)}{(\eLOUan - \eLOLan)}} .
  \label{eq:ratio2}
\end{align}
However, when using a reference readout, the squeezer characterization of \cref{sec:sqz-source} is convenient and sufficient. The above expressions are useful to check if the squeezing levels are changing between reference measurements and optical system measurements.

Using the normalized ADF-LO signal is advantageous because it simplifies calibrations. For instance, this normalization cancels scale factors from the beatnote signal, such as the transmitted audio field amplitude $\delta$, or the propagation and readout losses that constitute $\Lambda$. While cancelling these factors also allows minute duration drifts (i.e. drifts over the ADF scan) in generated squeezing levels $\delta$ and system losses $\Lambda$ to influence the beatnote measurement, these drifts are expected to be small. More importantly, the normalized beatnote signal is practical because it factors out frequency-dependent phase delays that the ADF picks up for technical reasons, from e.g. propagation delay and electronics. 
For our audio field measurements, we measure the normalized beatnote signal to remove such phase delays from our measurements of the system response.

Having derived how the quadrature observables relate to audio beatnote measurements, we can now calculate $m_q$ and $m_p$ for a given optical system $\Hr$ using Eq. \ref{eq:m_qp}.
The following analysis focuses on a filter cavity and a gravitational-wave interferometer with frequency-dependent squeezing, optical systems we would like to characterize using the ADF in the near future. 

\begin{figure*}
    \centering
    \includegraphics[width=\linewidth]{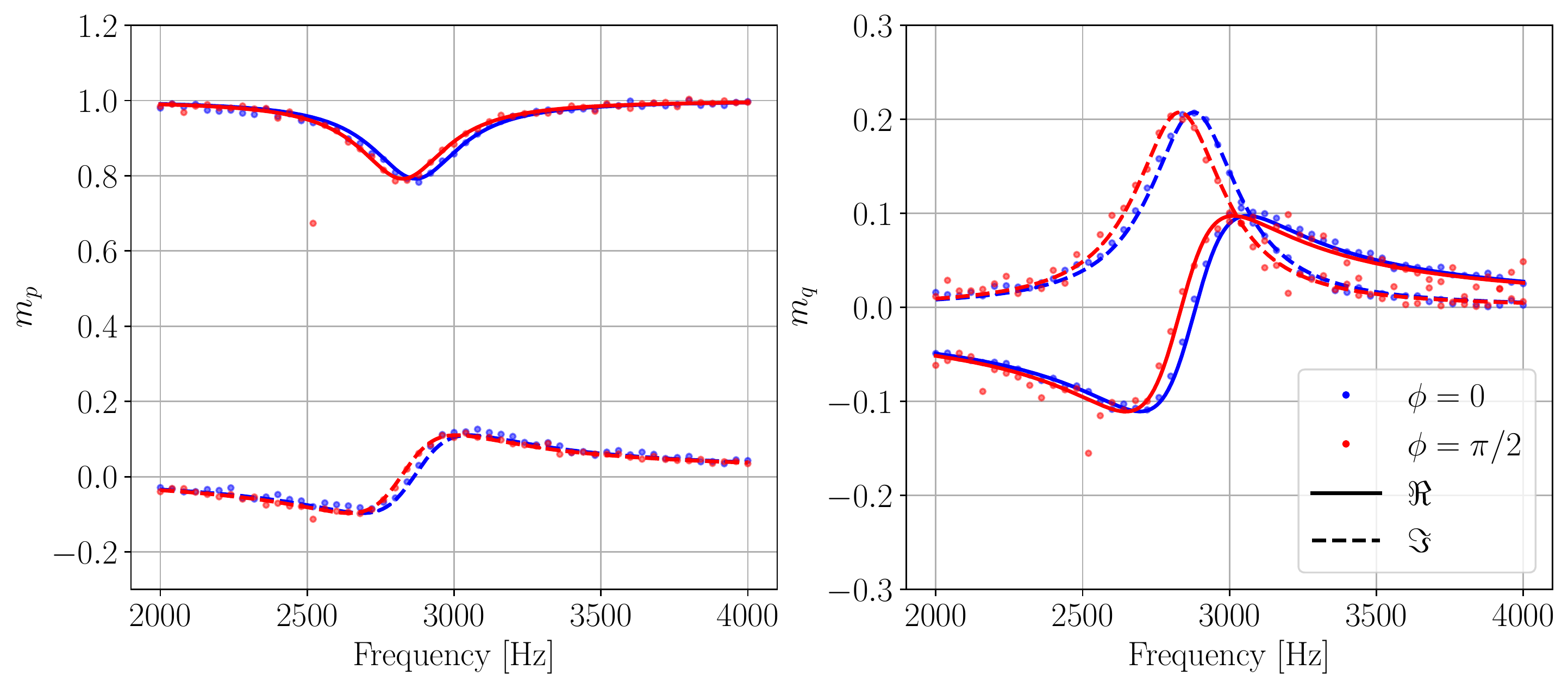}
    \caption{Characterization of a 16-m filter cavity using a sweep of the audio diagnostic field. Quadrature observables $m_p$ and $m_q$ are calculated from measurements of the normalized ADF-LO beatnote $\eLOno$ at two different squeezing angles, squeezing ($\angSQZ = 0$, blue) and anti-squeezing ($\angSQZ = \pi/2$, red), using Eqs. \ref{eq:mp_no} and \ref{eq:mq_no}. The normalized beatnote data is compared to the filter cavity model from Eq. \ref{eq:m_pq_fc}. The plot presents the experimental data, demodulated into real (solid) and imaginary (dashed) parts, along with the model curves fit to the data. The generated squeezing level measured by demodulating the ADF after the filter cavity (Eq. \ref{eq:ratio2}) was verfied against the squeezing level measured by demodulating the ADF directly after the squeezer (Eq. \ref{eq:e_gain}). 
    The fit parameters are given in Table \ref{tab:result_params}. }
    \label{fig:figure_measurements}
\end{figure*}

\subsection{Filter Cavity}
\label{sec:fc}
In the high finesse limit, the reflectivity of a filter cavity is \cite{Kwee-PRD14-DecoherenceDegradation,Whittle-PRD20-OptimalDetuning}   
%
\begin{equation}
    r_{\text{fc}}(\Omega) = \mm\Bigg(\frac{-\gamma + \lambda + \i(\Omega - \Delta \omegaFC)}{\gamma + \lambda + \i(\Omega - \Delta \omegaFC)}\Bigg)+(1-\mm),
    \label{eq:rp_fc}
\end{equation}
 expressed in terms of the \textit{coupler-limited} and \textit{loss-limited} HWHM bandwidths
\begin{equation}
  \label{eq:gammalambda}
  \gamma = \frac{c\Tin}{4\lengthFC},\ \lambda = \frac{c\loss}{4\lengthFC} ,
\end{equation}
where $\Tin$ is the input mirror transmissivity, $\loss$ is the filter cavity round trip loss, $\lengthFC$ is the cavity length, $\mm$ is the mode-matching efficiency from the squeezer to the filter cavity, and $\Delta\omegaFC$ is the filter cavity detuning from the carrier frequency.

In the quadrature basis, the filter cavity transfer matrix $\Hfc$ is given by 
\begin{align}
\Hfc &= \bA\begin{bmatrix}
\rp & 0 \\0 & \rmin
\end{bmatrix}\bAi ,\text{ using }
    &
      \begin{aligned}
\rp &= r_{\text{fc}}(\Omega)
\\
\rmin &= r^*_{\text{fc}}(-\Omega)
\end{aligned}
\end{align}
where $\rp$ and $\rmin$ are the complex filter cavity reflectivity at frequencies corresponding to the upper and lower sideband, respectively.
For $\Hfc$, the quadrature observables calculated using Eq. \ref{eq:m_qp} are given by
\begin{align}
 m_p = \frac{1}{2}(\rp+\rmin)
    \label{eq:m_pq_fc}
    && m_q = -\frac{i}{2}(\rp-\rmin) . \end{align}
Combined with Eqs. \ref{eq:mp_no}-\ref{eq:mq_no}, our analysis connects measurements of the normalized ADF-LO beatnote $\eLOno$ to both the optical response of the filter cavity $\Hfc$, and squeezed state propagation through the cavity.

\begin{table}[b]
\centering
    \begin{ruledtabular}
    \begin{tabular}{l|c}
        \textbf{Parameter}&Value \\
        \hline
        
        &\textit{Independently} \\
        &\textit{Measured} \\
        \hline
        OPO $M_1$ Reflectivity ($r_1^2$)&0.875\\
        OPO $M_2$ Reflectivity ($r_2^2$)&0.9985\\
        Filter cavity length ($\lengthFC$)&16 m \\
        Filter cavity input mirror & 63.7 ppm\\
         transmissivity ($\Tin$)& \\
        \hline
        &\textit{Estimated using} \\
        &\textit{ the ADF} \\
        \hline
        Generated Squeezing in OPO & 5.56 dB\\
        Filter cavity round-trip loss ($\loss$)&181 ppm \\
        Filter cavity mode matching ($\mm$)&0.799\\
        Filter cavity detuning ($\Delta\omegaFC/2\pi$) &  2879 Hz ($\angSQZ = 0$)\\
        &2830 Hz ($\angSQZ = \pi/2$)\\
        
    \end{tabular}
    \end{ruledtabular}
    \caption{Experimentally determined parameters of the OPO and 16 m filter cavity used for frequency-dependent squeezing.}
    \label{tab:result_params}
\end{table}

\textbf{Experimental results.} Figure \ref{fig:figure_measurements} shows the use of audio field diagnostics of our 16-m filter cavity, and its preparation of a frequency-dependent squeezed state.

To characterize the filter cavity at 3 kHz detuning, the ADF frequency $f$ was swept from 2 kHz to 4 kHz. $m_p$ and $m_q$ are calculated from the normalized ADF-LO beatnote signals $\eLOno$ (Eq. \ref{eq:eLO_norm}), obtained by balanced homodyne detection of the audio field after passage through the 16-m filter cavity. To normalize the ADF sweep and isolate the filter cavity response, the audio sweep was performed with the filter cavity locked near-resonance with the squeezed field, and then again with the cavity off-resonance; the on- and off- resonance responses were divided to yield the normalized beatnote signals $\eLOno$.
$m_p$ and $m_q$ were then calculated from measurements of $\eLOno$ using Eqs. \ref{eq:mp_no} and \ref{eq:mq_no}. 

Table \ref{tab:result_params} summarizes the filter cavity parameters extracted from fits to these audio sweep measurements. 
Data obtained from the normalized LO beatnote were converted to $m_p$ and $m_q$ using Eqs. \ref{eq:mp_no}, \ref{eq:mq_no} and fit to the filter cavity model described in Eq. \ref{eq:m_pq_fc} in order to estimate the cavity detuning $\Delta\omegaFC$, round trip loss $\loss$, and mode-matching $\mm$ with the squeezed vacuum field. The squeezing level was also measured using the normalized ADF-LO beatnote $\eLOno$ (Eq. \ref{eq:ratio2}). The precise filter cavity detuning has percent-level variations between squeezing and anti-squeezing, due to technical challenges in stabilizing the filter cavity length at kilohertz detunings \cite{McCuller-PRL20-FrequencyDependentSqueezing}. Relevant parameters measured independently without the audio field include the filter cavity's input mirror transmissivity $\Tin$, and the reflectivity of mirror $M_1$ in the OPO $r_1$. 



\begin{figure}
    \centering
    \includegraphics[width=\linewidth] {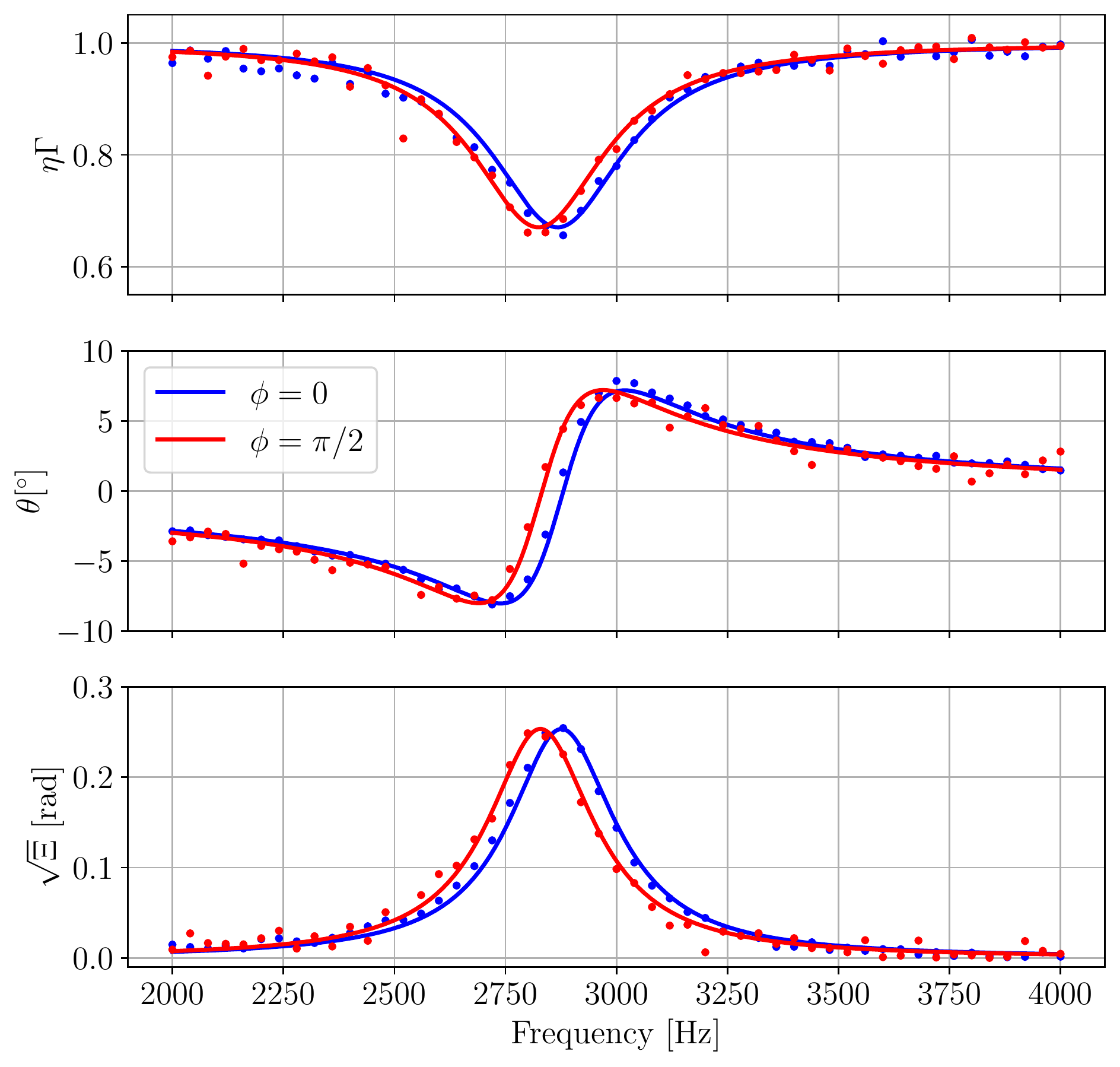}
    \caption{ Filter cavity loss, rotation, and dephasing calculated from $m_p$ and $m_q$ (from Fig. \ref{fig:figure_measurements}) using Eqs. \ref{eq:fdloss}, \ref{eq:fdrot} and \ref{eq:fddep} plotted along with model curves (solid). The top plot corresponds to the squeezing efficiency $\fdloss$ multiplied by the noise gain $\Gamma$ which is 1 for the filter cavity. The middle plot shows the squeezing rotation $\fdrot$ in degrees. The bottom plot contains the square root of the frequency-dependent dephasing $\fddep$. $\sqrt{\fddep}$ has a similar effect on squeezing as phase noise $\Delta\phi$ with the same RMS value, and therefore it has been represented in units of radians. Blue corresponds to squeezing while red corresponds to anti-squeezing.}
    \label{fig:fdloss}
\end{figure}

In Fig. \ref{fig:fdloss}, the squeezing degradation is calculated from the $m_p$ and $m_q$ data using Eqs. \ref{eq:fdloss}, \ref{eq:fdrot} and \ref{eq:fddep} to determine the frequency-dependent squeezing efficiency $\fdloss\Gamma$, rotation $\fdrot$, and dephasing $\fddep$ introduced by the filter cavity. 
From this plot it is easy to see that the high round trip loss in the filter cavity limits the squeezing rotation to less than $10^\circ$ while adding considerable dephasing and squeezing loss. As the these specific degradation mechanisms affect the frequency-dependent squeezing spectra in degenerate ways, it is difficult to distinguish between them using photodetector noise spectrum measurements. Data for the above results were obtained much faster and more reliably than prior means to characterise a filter cavity using frequency-dependent squeezing in photodetector quantum noise spectra \cite{Komori-PRD20-DemonstrationAmplitude}. 

%








\begin{figure*}
    \centering
    \includegraphics[width=\linewidth]{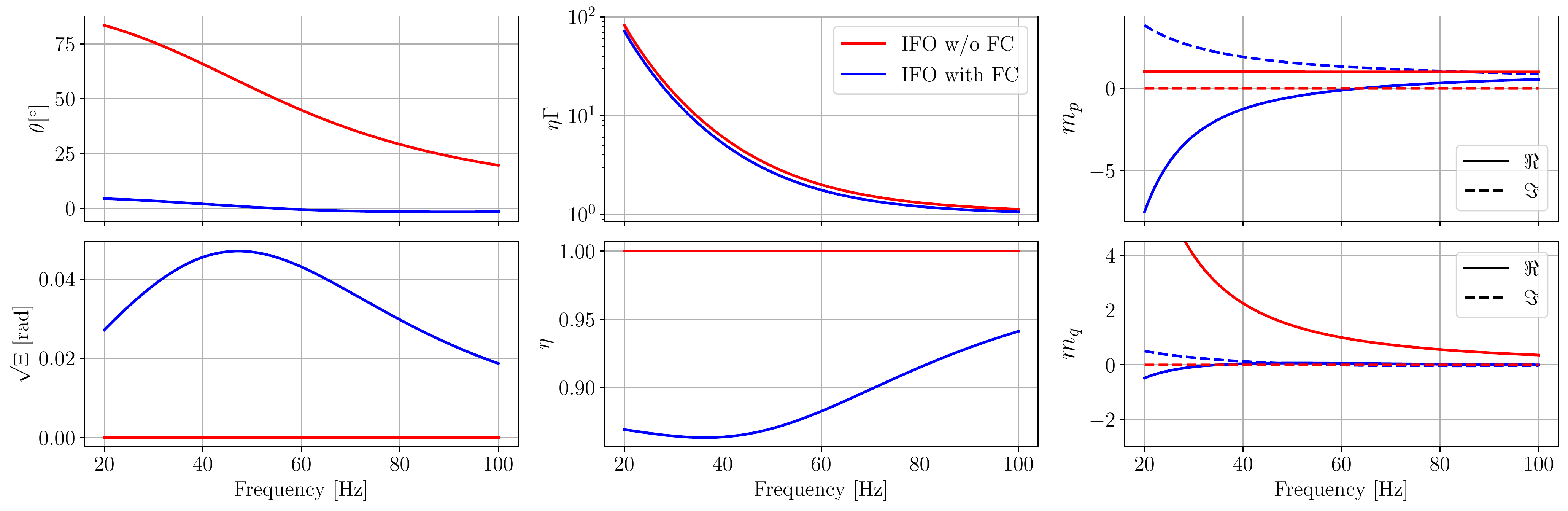}
    \caption{Simulation of quadrature observables and squeezing degradation metrics for an ideal interferometer with (blue) and without (red) a filter cavity. The left curves correspond to the rotation $\fdrot$ and dephasing $\fddep$ ($\sqrt{\fddep}$ has been represented as equivalent RMS phase noise in radians) of the two configurations. The center plots contain the squeezing efficiency $\fdloss$ along with the optomechanical gain from the interferometer $\Gamma$. While the loss and gain cannot be measured independently, we may assume an ideal inteferometer without a filter cavity should have $\fdloss=1$. We can back out the loss of the inteferometer-filter cavity  by normalizing to the gain of the lossless interferometer.  The right plots show the quadrature observables $m_p$, $m_q$ which have been generated using Eq. \ref{eq:m_pq_IFO}. Eqs. \ref{eq:fdloss}-\ref{eq:fddep} are used to convert the quadrature observables into squeezing metrics. It is assumed that the interferometer reaches the standard quantum limit at a frequency of $\Omega_\text{SQL} = 2\pi\cdot 60$ Hz. 
    Simulation parameters, representing design specifications for frequency-dependent squeezing in LIGO \cite{AplusDesign}, assume round trip filter cavity losses of 60 ppm, input mirror transmissivity of 1200 ppm, cavity detuning of 43 Hz, and mode matching efficiency of 0.99. }
    \label{fig:figure_sim}
\end{figure*}

\subsection{Interferometer}
\label{sec:ifo}

We now consider the response of an ideal, dual-recycled Fabry-Perot interferometer, such as LIGO. The following calculations assume on-resonance operation in a lossless interferometer, which has no mode mismatch with the injected squeezed beam. Such an interferometer can be represented by the following two-photon matrix,
\begin{equation}
    \Hifo \simeq \begin{bmatrix}
    \IFOr&0\\-\IFOk&\IFOr
    \end{bmatrix}
\end{equation}
where $\IFOr$ is related to the signal bandwidth $\IFOg$ of the interferometer
\begin{equation}
    \IFOr \simeq \frac{\IFOg -i\Omega}{\IFOg +i\Omega} ,
\end{equation}
and $\IFOk$ is the interaction strength of the interferometer which is related to the radiation pressure noise caused by optomechanical coupling. It is related to the standard quantum limit (SQL) frequency $\OmegaSQL$, where quantum radiation pressure noise has the same magnitude as shot noise, as 

\begin{equation}
 \IFOk = -\frac{\OmegaSQL^2}{\Omega^2}\Big(\frac{\IFOg}{\IFOg+i\Omega}\Big)^2.
\end{equation}

The transfer matrix for propagation through a filter cavity followed by an ideal interferometer is 
\begin{equation}
 \Hr = \Hifo\Hfc ,
\end{equation}
from which Eq. \ref{eq:m_qp} yields the quadrature observables 
\begin{align}
    m_p = \frac{\IFOr}{2}(\rp+\rmin)-\frac{i\IFOk}{2}(\rp-\rmin) \\
    m_q = -\frac{i\IFOr}{2}(\rp-\rmin)-\frac{\IFOk}{2}(\rp+\rmin).
    \label{eq:m_pq_IFO}
\end{align}

The introduction of a filter cavity enables us to inject frequency-dependent squeezing into the inteferometer. The squeezing rotation produced by the filter cavity aims to counteract the rotation due to the interferometer's optomechanical coupling. This is important at frequencies around and below the SQL of the interferometer. For these frequencies well within the signal bandwidth, i.e. $\Omega\ll\IFOg$, we can assume that $\IFOr\approx1$ and $\IFOk \approx -\OmegaSQL^2/\Omega^2$. 

Fig. \ref{fig:figure_sim} contains simulation results of quadrature observables and squeezing metrics for an inteferometer with and without a filter cavity. We observe that the interferometer (red curve), through its optomechanics, produces a squeezing rotation of around 90$^\circ$ at low frequencies, which is equivalent to rotating squeezing into anti-squeezing. The addition of an optimally detuned \cite{Whittle-PRD20-OptimalDetuning} filter cavity (blue curve) seeks to reverse this rotation. However, the filter cavity also introduces loss and dephasing, which can degrade the squeezing measured at the readout. The parameters of the simulation have been chosen to be representative of the frequency-dependent squeezing upgrade to LIGO \cite{AplusDesign}. The filter cavity has been designed in order to optimize squeezing rotation while keeping squeezing degradation to a minimum. The ADF can help diagnose these effects, which would be difficult to measure otherwise, and consequently inform operational choices for filter cavities in interferometers. 

With and without a filter cavity, the ADF directly measures the interferometer noise gain and efficiency $\fdloss\Gamma$. This parameter can be used to determine the true local oscillator angle of the readout by its effect on the noise gain, as was done using an involved squeezing measurement \cite{McCuller-PRD21-LIGOQuantum}. We anticipate the ADF can provide additional diagnostics while commissioning a balanced homodyne detection upgrade in gravitational-wave interferometers. Such an upgrade enables freely changing the local oscillator angle, but only implicitly knowing the angle from a control system error point and the calibrated signal sensitivity. The noise gain is useful to know precisely, as it scales the magnitude of certain classical noises in the interferometer, such as backscatter.

\textbf{Time-resolved fluctuations}.
In addition to jointly characterizing the interferometer and filter cavity, the ADF also enables a time-resolved view of how large-RMS interferometer motions will degrade squeezing. In particular, insufficiently controlled motions of the interferometer and filter cavities will cause drifting squeezing rotations inside the resonance of the drifting cavities, leading to frequency-dependent phase noise. This form of degradation is difficult to resolve using squeezing alone, because squeezing spectrum measurements require integrating for longer than drift time-scales and at multiple squeezing levels. The ADF can probe for changing squeezed state rotation at specific frequencies.

The first such example is using the ADF above the interferometer arm bandwidth $\gamma_\Arm\approx 2\pi{\cdot}430$ Hz (for LIGO) but within the signal cavity bandwidth $\gamma_\Src\approx 2\pi{\cdot}80$ kHz. At these frequencies, the signal extraction cavity may have residual motion as its length $L_\src$ drifts, which causes the squeezed state to rotate. The frequency dependence of that rotation, given length fluctuations, is given by Eq. 69 of \cite{McCuller-PRD21-LIGOQuantum}, which can be expressed 
\begin{align}
  \frac{\mathrm{d} \theta_{\text{IFO}}(\Omega)}{\mathrm{d} L_\src}
  &\approx
    \frac{-8k}{T_\srm}
    \left(
    \frac{\gamma^2_\Src}{\gamma_\Src^2 + \Omega^2}
    -
    \frac{\gamma^2_\Arm}{\gamma_\Arm^2 + \Omega^2}
    \right),
    \label{eq:SRM_dTheta_dL}
\end{align}
where $k$ is the wavenumber of the carrier light. Similarly, length changes of the filter cavity will cause its rotation to change. For a lossless filter cavity at its optimal detuning, the sensitivity of the squeezed state rotation to length changes is given by the derivative of Eq. 18 in \cite{Kwee-PRD14-DecoherenceDegradation}, which can be written
\begin{align}
  \frac{\mathrm{d} \theta_{\text{FC}}(\Omega)}{\mathrm{d} L_{\text{fc}}}
  &\approx
    \frac{-8k}{T_{\text{fc}}}
    \left(
    \frac{\gamma^2_{\text{fc}}\Omega^2}{\Omega^4 + 4 \gamma_{\text{fc}}^4}
    \right) \le \frac{-4k}{T_{\text{fc}}}.
    \label{eq:FC_dTheta_dL}
\end{align}
Here $\gamma_{\text{fc}}$ is the HWHM bandwidth of the filter cavity, optimal at $\gamma_{\text{fc}} \approx \Omega_{\text{SQL}} / \sqrt{2}$, and the cavity is detuned by its bandwidth $\gamma_{\text{fc}}$ to cause a $90^\circ$ rotation. This expression indicates the high sensitivity of the detuned filter cavity to length noise, due to its small input transmissivity required to create its small bandwidth.

For the interferometer phase drift measurement, where $\gamma_\Arm < \Omega < \gamma_\Src$, the normalized beatnote measurement can be used. There, the interferometer should not be changing the sideband balancing, so $\eLOUn = \eLOLn$.
Under that condition, it is possible to measure the fluctuations in the effective squeezing angle caused by these changes using the methods described in Sec. \ref{sec:sqz-source}. 

For a lossy, detuned filter cavity with or without the interferometer in series, $\eLOUn \neq \eLOLn$, so 
the methods of \cref{sec:opt-sys} are required to calculate $\theta$. This poses a problem for time resolved measurements, as both the upper and lower ADF cannot be simultaneously driven. In this case, the effects of length fluctuations of a specific system model on the ADF is required to relate independent upper and lower ADF measurements along with measurements during squeezing and anti-squeezing.


\section{Conclusions}
\label{sec:conclusion}

The upcoming upgrades to Advanced LIGO and Advanced Virgo, and plans for future gravitational-wave detectors, are contingent on high levels of squeezing. Measured squeezing levels are sensitive to degradation mechanisms such as loss and phase noise. Previous methods to measure and characterize these mechanisms have generally had to use measurements of squeezed noise spectra, which require extensive setup and long averaging times.

This work provides an alternative characterization method which uses an audio sideband injected into the squeezer, that co-propagates with the squeezed vacuum beam. This method can be used to perform in-situ measurements of the squeezed vacuum source and rapidly characterize optical systems such as filter cavities and interferometers. As a proof of concept, this work presents experimental results from using the ADF in order to characterize a 16 m filter cavity used for frequency-dependent squeezing. 

We plan to use this method in the near future to characterize LIGO during the observing run O4, where frequency-dependent squeezing will be employed for the first time to reduce quantum radiation pressure noise. In addition to commissioning the filter cavity, this analysis of audio field diagnostics may allow us to reliably characterise and understand the optomechanics of the interferometer and specific effects such as loss in the signal recycling and arm cavities. This technique will advance our understanding of squeezing generation and degradation for future iterations of gravitational-wave detectors, and help in breaking the limits of quantum noise reduction in current interferometers.

\section*{Acknowledgements}
This material is based upon work supported by NSF's LIGO Laboratory which is a major facility fully funded by the National Science Foundation, operating under Cooperative Agreement No. PHY-1764464. Advanced LIGO was built under Grant No. PHY-0823459. The A+ upgrade to Advanced LIGO is funded under Grant No. PHY-1834382. WJ is supported by the MathWorks, Inc. The authors gratefully acknowledge important comments on this manuscript from Sheila Dwyer and Kevin Kuns, and discussions within the LIGO Scientific Collaboration working groups during the development of this project. 

\appendix

\section{Squeezer Model}
\label{section:squeezer}
The squeezer model used in this paper corresponds to the bowtie cavity optical parametric resonator (OPO) shown in Figure \ref{fig:figure_exp}. Using the sideband basis, the single pass matrix of the squeezer crystal is given by 
\begin{equation}
\Smat{S} = \begin{bmatrix}
\SQZcosh&&e^{-i2\angPUMP}\SQZsinh\\
e^{i2\angPUMP} \SQZsinh&&\SQZcosh
\end{bmatrix}
\label{eq:OPO}
\end{equation}
where
$\SQZint$ is the single pass squeeze factor of the squeezer's nonlinear crystal, and $2\angPUMP$ is the squeezer pump phase. The model assumes that the squeezer is in the pump-undepleted regime, i.e the decrease in squeezer pump power (which is related to $\SQZint$) due to non-linear down conversion is negligible. Converting the above expression to the quadrature picture shows us that the squeezer matrix is diagonal and is given by

\begin{equation}
\Smat{S} = \bA^\dagger \mathbf{R}_\psi
\begin{bmatrix}
    e^z & 0\\
    0 & e^{-z}
    \end{bmatrix}
    \mathbf{R}_\psi^\dagger \bA
    \label{eq:OPOq}
\end{equation}
where the rotation matrix $\mathbf{R}_\psi$ is defined conventionally as
\begin{equation}
    \mathbf{R}_\psi = 
\begin{bmatrix}
    \cos \psi & -\sin \psi \\
    \sin \psi & \cos \psi
\end{bmatrix} .
\end{equation}

\subsection{ADF Transfer Matrix}

For the following calculations, it is assumed that the OPO is lossless, and the reflectivity of $M_3$ and $M_4$ is 1. It is also assumed that the reflectivity of $M_2$ is very close to 1 ($r_2\approx 1$). 

The cavity round-trip phase shift, in matrix form, is given by

\begin{equation}
\thcav = \begin{bmatrix}
e^{-i\angDET}&&0\\
0&&e^{i\angDET}
\end{bmatrix}
\end{equation}
where
the cavity detuning angle $\angDET$ is 
\begin{equation}
\angDET = \frac{2\pi f}{f_{\text{FSR}}},
\end{equation}
$f$ is the sideband frequency, and $f_{\text{FSR}}$ is the free spectral range of the OPO cavity.

In our setup, auxiliary fields such as the CLF and ADF are injected into the OPO via the transmission mirror $M_2$. The transfer matrix from the transmission from $M_2$ to $M_1$ is given by 

\begin{equation}
    \Hopo = t_1t_2({\mathbf{I}-r_1\mathbf{\thcav\Smat{S}}})^{-1}
\end{equation}

\begin{equation} \label{eq: ADFmatrix}
\mathbf{\Hopo} = 
    t_1t_2\dfrac{
    \begin{bmatrix}
1-e^{i\angDET}r_1\SQZcosh&&r_1e^{-i(2\angPUMP+\angDET)}\SQZsinh\\
r_1e^{i(2\angPUMP+\angDET)}\SQZsinh
&&1-e^{-i\angDET}r_1\SQZcosh
\end{bmatrix}
}{ r_1^2 -2 r_1\SQZcosh\cos\angDET +1} ,
\end{equation}
up to a global phase, which has been omitted.

Note that this expression is only valid for the case where a beam injected via $M_2$ exits the OPO through $M_1$ before passing through the squeezer crystal. Alternate OPO configurations are explored in Appendix \ref{sec:alt_opo}.

\subsection{Effective Squeeze Factor of OPO}

The OPO pump field and quantum vacuum, are injected via $M_1$, and the squeezed vacuum is obtained on reflection at $M_1$. For our specific OPO configuration, the effective OPO reflection matrix at $M_1$ is given by 
\begin{equation}
    \Hopor = r_1\mathbf{I} - t_1^2\thcav\Smat{S}({\mathbf{I}-r_1\mathbf{\thcav\Smat{S}}})^{-1} .
\end{equation}

The cavity is operated at resonance for the squeezed vacuum carrier frequency, i.e. $\angDET=0$. Solving for $\Hopor$ (and comparing with Eq. \ref{eq:OPOq}) produces an modified squeezer matrix.  squeeze factor

 \begin{equation} 
   \Hopor =\bA^\dagger \mathbf{R}_\psi
    \begin{bmatrix}
    \frac{t_1^2 e^z}{1-r_1  e^z}-r_1 && 0\\
    0 &&  \frac{t_1^2  e^{-z}}{1-r_1  e^{-z}}-r_1
    \end{bmatrix}
      \mathbf{R}^\dagger_\psi  \bA 
      \label{eq:Ho_quadrature}
\end{equation}  
\begin{equation} 
    = \bA^\dagger \mathbf{R}_\psi
    \begin{bmatrix}
    e^{Z} & 0\\
    0 & e^{-Z}
    \end{bmatrix}
      \mathbf{R}^\dagger_\psi  \bA
      \label{eq:z_eff_exp}
\end{equation}
where $Z$ is the effective squeeze factor. Relating \cref{eq:Ho_quadrature} and \cref{eq:z_eff_exp} then gives Eq. \ref{eq:zZe}

\section{Comparison with Previous Literature}
\label{section:NLG}
The transmission ratio $G$ (Eq. \ref{eq:GR}) of an injected near-carrier field is similar to the transmission parametric gain factors sometimes referred as the ``nonlinear gain,'' $\overline{g}$, in existing and widely used calibration schemes \cite{Lam-JOBQSO99-OptimizationTransfer, Bowen-03-ExperimentsQuantum, Zhang-PRA03-QuantumTeleportation, Aoki-OEO06-Squeezing946nm, Takeno-OEO07-ObservationDB}. Specifically, $\overline{g}$ refers to the gain in power of a transmitted seed field at the carrier frequency, with respect to no non-linearity. The main difference between the two is that $G$ is a ratio of the maximum to minimum quadrature gain in field units while $\overline{g}$ is a ratio of the maximum quadrature power to the ``gain-free'' $z=0$ transmission power. The nonlinear gain measurement using a seed beam is defined in the degenerate limit $f=0$, but it can expressed in our conventions as
\begin{align}
  \overline{g} &= \frac{\max(|\ADFU+\ADFL|^2)}{\lim_{e^z \rightarrow 1} |\ADFU+\ADFL|^2} = \left|\frac{1 - r_1}{1 - r_1e^{z}}  \right|^2 .
  \label{eq:NLG}
\end{align}

Note that the expression for power includes both the upper and lower sidebands. This is because, in the sideband picture, a carrier field is represented as a sum of both upper and lower sidebands at $f=0$. Typical usage of $\overline{g}$ relates the transmission gain to the squeezing level using the Collett-Gardiner model \cite{Collett-PRA84-SqueezingIntracavity} for the OPO. This model uses a Hamiltonian formulation that linearizes the internal squeezing operation within the cavity. The frequency-domain input-output model of App.~\ref{section:squeezer} does not linearize the internal squeezing, which is why \cref{eq:NLG} is not exactly the same as in the references. For finite finesse OPO's with the parameters used in this paper, our model provides a few percent correction in its estimate of the squeezing level $e^{-2Z}$ over the linearized model, but requires formulas that depend on the specific layout of the OPO.  In the high finesse $r_1\rightarrow 1$ limit, the two models are consistent.

A practical advantage of using the ADF transmission ratio $G$ as opposed to the non-linear gain $\overline{g}$ is that measurements of the latter usually require a significant amount of carrier power. This can lead to pump-depletion and deviation from the squeezer model. On the other hand, accurate measurements of $G$ do not require a high ADF power, as the ADF-LO beatnote can be made arbitrarily large by increasing the LO power. 

\section{Alternate OPO Configurations}
\label{sec:alt_opo}
\begin{figure}
    \centering
    \includegraphics[width = \linewidth]{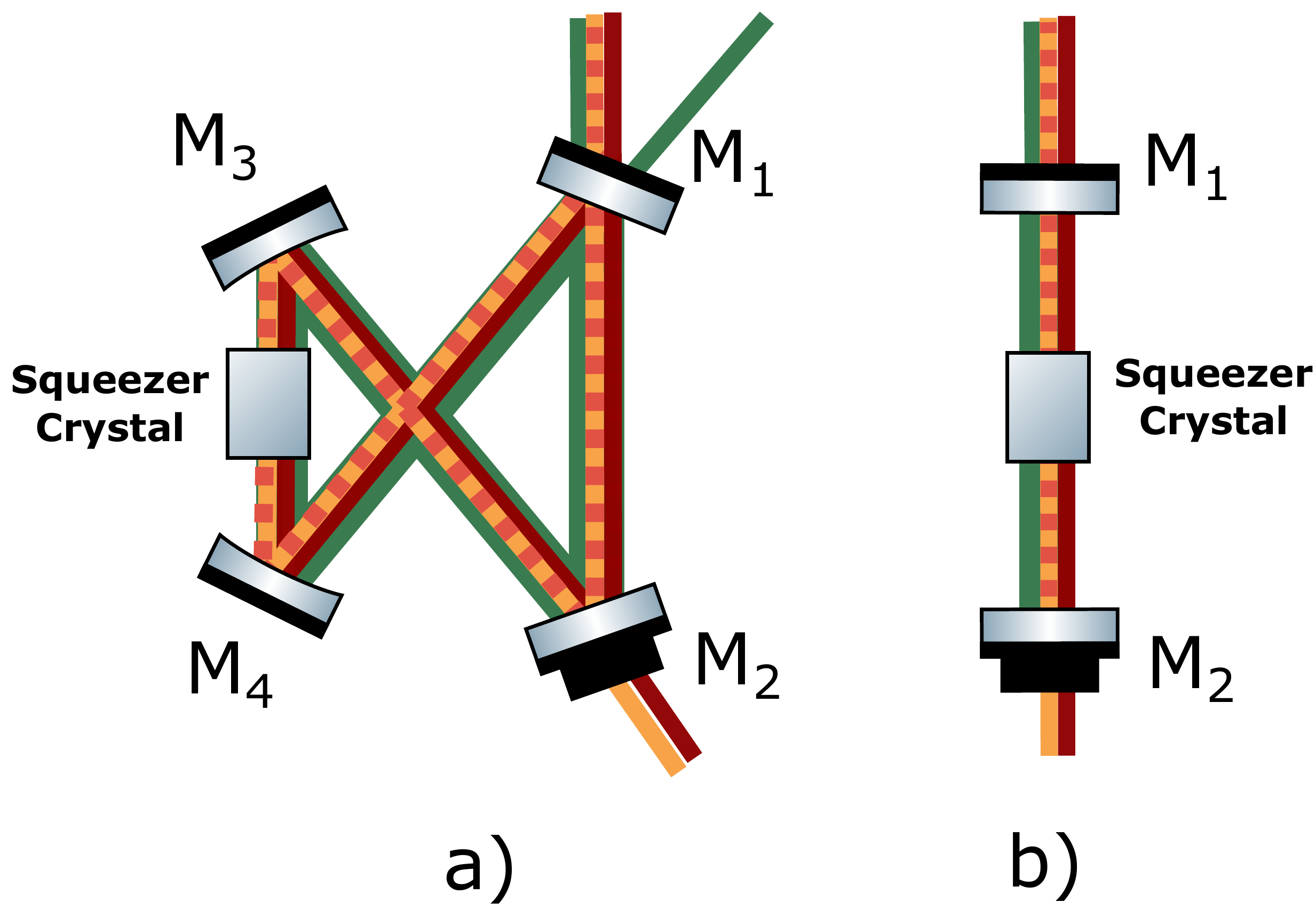}
    \caption{Alternate OPO configurations. a) Alternate bowtie, b) Linear cavity. }
    \label{fig:alt_OPO}
\end{figure}

In this section, we explore how the alternate OPO configurations shown in Fig. \ref{fig:alt_OPO} change the ADF transfer matrix through the cavity.
\subsection{Alternate bowtie }
In the alternate bowtie configuration shown in Fig. \ref{fig:alt_OPO} a), a beam injected via $M_2$ passes through the crystal before exiting the OPO through $M_1$, as opposed to exiting the crystal first as it does in the experimental setup of this paper. In this case, the transfer matrix is modified to 
\begin{equation}
    \Hopo = t_1t_2\Smat{S}({\mathbf{I}-r_1\mathbf{\thcav\Smat{S}}})^{-1}
\end{equation}
\begin{equation}
\mathbf{\Hopo} = 
    t_1t_2\dfrac{
    \begin{bmatrix}
e^{i\angDET}\SQZcosh-r_1&&e^{-i(2\angPUMP+\angDET)}\SQZsinh\\
e^{i(2\angPUMP+\angDET)}\SQZsinh
&&e^{-i\angDET}\SQZcosh-r_1
\end{bmatrix}
}{ r_1^2 -2 r_1\SQZcosh\cos\angDET +1} .
\end{equation}
The ADF exiting the OPO then becomes

\begin{equation}
  \ADFU = \aADF \dfrac{t_1t_2e^{-i {\angADF}}}{ r_1^2 -2 r_1\SQZcosh+1}\begin{bmatrix}
   \SQZcosh -r_1\\e^{i2\angPUMP}\SQZsinh
    \end{bmatrix} .
\end{equation}  
This can also be written as Eq. \ref{eq:ADF_simp} with the modified parameters 
\begin{align}
  \alpha &= \SQZcosh-r_1
  & 
    \beta &=  \SQZsinh ,
\end{align}
which modifies Eq. \ref{eq:GR} for this alternate configuration to
\begin{equation}
    G = \frac{e^\SQZint-r_1}{e^{-\SQZint}-r_1}.
\end{equation}

The effective squeezing $Z$ of this bowtie configuration is the same as the original bowtie (Eq. \ref{eq:zZe}), as a roundtrip in the cavity is identical in both configurations. 

\subsection{Linear Cavity}
In the linear configuration shown in Fig. \ref{fig:alt_OPO} b), we consider a transmitted beam that makes a double pass through the crystal during every roundtrip, and then makes a single pass through the crystal before exiting the OPO. The transfer matrix is then modified to

\begin{equation}
    \Hopo = t_1t_2\Smat{S}({\mathbf{I}-r_1\mathbf{\thcav\Smat{S}^2}})^{-1}
\end{equation}

\begin{equation}
\mathbf{\Hopo} = 
    t_1t_2\dfrac{
    \begin{bmatrix}
\SQZcosh(1-e^{i\angDET}r_1)
&&e^{-i2\angPUMP}\SQZsinh(1+e^{-i\angDET}r_1)\\
e^{i2\angPUMP}\SQZsinh(1+e^{i\angDET}r_1)
&&\SQZcosh(1-e^{-i\angDET}r_1)
\end{bmatrix}
}{ r_1^2 -2 r_1\SQZcosht\cos\angDET +1} ,
\end{equation}
such that injecting upper audio sideband into the OPO produces the transmitted audio field,
\begin{equation}
  \ADFU = \aADF \dfrac{t_1t_2e^{-i{\angADF}}}{ r_1^2 -2 r_1\SQZcosht+1}\begin{bmatrix}
    \SQZcosh(1-r_1)\\e^{i2\angPUMP}\SQZsinh(1-r_1)
    \end{bmatrix} .
\end{equation}  

In a linear cavity, a cavity round-trip involves passing the squeezer crystal twice. We can replace this with an effective single pass through a crystal with double the squeeze factor and scale $z$ down to $z/2$. Similar to the alternate bowtie case, we can write this in terms of Eq. \ref{eq:ADF_simp}  

\begin{align}
  \alpha &= (1-r_1)\SQZcoshh 
  & 
    \beta &=  (1+r_1)\SQZsinhh ,
\end{align}
modifying Eq. \ref{eq:GR} for the linear cavity to
\begin{equation}
    G = \frac{e^\SQZint-r_1}{1-e^{\SQZint}r_1}.
\end{equation}
The effective squeeze factor of the crystal remains unchanged with respect to other configurations (Eq. \ref{eq:zZe}).

\nocite{apsrev42Control}
\bibliographystyle{apsrev4-2-trunc}
\bibliography{paper}

\end{document}